\documentclass[a4paper,10pt,oneside]{article}
\usepackage{epsfig}
\usepackage{hangcaption}
\usepackage{ngerman}
\usepackage{graphicx}
\usepackage[bf]{caption}
\usepackage[dvips]{hyperref}
\usepackage{fancyhdr}
\renewcommand{\headrulewidth}{0.4pt}

\newcommand{\GeV}{\mbox{$\mathrm{GeV}$}}
\newcommand{\TeV}{\mbox{$\mathrm{TeV}$}}
\newcommand{\EGeV}{\mbox{$E(\mathrm{GeV})$}}
\newcommand{\roots}{\mbox{$\sqrt{s}$}}
\newcommand{\qq}{\mbox{$\mathrm{q\overline{q}}$}}

\newcommand{\qqp}{\mbox{$\mathrm{q'\overline{q}}$}}

\newcommand{\uds}{\mbox{$\mathrm{uds}$}}

\newcommand{\Zzero}{\mbox{${\mathrm{Z}^{0}}$}}
\newcommand{\Zuds}{\mbox{\Zzero$\rightarrow$\uds}}

\newcommand{\RMS}{\mathrm{RMS}_{90}}

\newcommand{\ILC}{\mbox{ILC}}

\newcommand{\LDC}{\mbox{LDC}}
\newcommand{\LCIO}{\mbox{\tt LCIO}}

\newcommand{\GEANT}{\mbox{\tt GEANT4}}
\newcommand{\MOKKA}{\mbox{\tt Mokka}}
\newcommand{\MARLIN}{\mbox{\tt Marlin}}
\newcommand{\MARLINRECO}{\mbox{\tt MarlinReco}}

\newcommand{\PFA}{\mbox{PFA}}
\newcommand{\WOLF}{\mbox{Wolf}}
\newcommand{\PANDORAPFA}{\mbox{\sc PandoraPFA}}

\newcommand{\NIMA}[3] {Nucl.\ Instr.\ and Meth.\ \textbf{A#1} (#2) #3}
\def\etal{\mbox{{\it et al.}}}

\hypersetup{pdfhighlight=/O,pdfpagemode=None,pdfstartview=XYZ,pdftitle={Track-Based Particle Flow},pdfauthor={O.~Wendt},pdfsubject={Track-Based Particle Flow},pdfkeywords={particle~flow,~event~reconstruction,~calorimetry}}

\title{Track-Based Particle Flow\textbf{\footnote{to appear in Proceedings of LCWS07, Hamburg, Germany, June 2007}}}
\author{O.~Wendt$^{1,2}$}

\date{}

\begin{document}
\maketitle

\vspace{-5ex}

\begin{table}[h]
\centering	
\begin{tabular}{r@{ }l}
$^{1}$ & DESY, Notkestrasse~85, 22607~Hamburg, Germany\\
$^{2}$ & University of Hamburg - Department of Physics,\\ 
       &  Luruper~Chaussee~149, 22761~Hamburg, Germany \\
       &
\end{tabular}
\end{table}

\renewcommand{\headrulewidth}{0pt}

\fancyhead{} 

\begin{abstract}
\noindent One of the most important aspects of detector development for the
\ILC\, is a good jet energy resolution $\sigma_E/E$. To achieve the
goal of high precision measurements $\sigma_E/E = 0.30/\sqrt{\EGeV}$
is proposed. The particle flow approach together with highly granular
calorimeters is able to reach this goal. This paper presents a new
particle flow algorithm, called Track-Based particle flow, and shows
first performance results for 45~\GeV\,jets based on full detector
simulation of the Tesla~TDR detector model.\\ \\
PACS~numbers:~07.05.Kf,~29.40.Vj,~29.85.+c
\end{abstract}

\section{Introduction}
The International Linear Collider~(\ILC) will provide the potential
for high precision measurements at center-of-mass energies between
several hundred \GeV\, and one \TeV. Many interesting physics
processes in this regime will be composed of multi-jet final states
originating from hadronic decays of heavy gauge bosons. To reach the
goal of high precision measurements it is suggested to achieve a mass
resolution for $\mathrm{W}\rightarrow\qqp$ and
$\mathrm{Z}\rightarrow\qq$ decays which is comparable to their
widths. This leads to a jet energy resolution of $\sigma_E/E =
0.30/\sqrt{\EGeV}$ considering the typical di-jet energies ranging
from $100$ to $400$~\GeV. Studies based on full detector simulation
have shown that particle~flow~algorithms~(\PFA)\, are able to reach
this goal~\cite{teslatdr,Thomson:2007xb}. The basic concept of any
\PFA\, is to reconstruct the four-momenta of all visible particles in
an event. The four-momenta of charged particles are measured in the
tracking detectors, while the energy of photons and neutral hadrons is
obtained from the calorimeters. The accuracy of momentum measurement
in the tracking systems is by orders of magnitude better than the
accuracy of energy measurement in the calorimeters. This leads to a
theoretical limit on the jet energy resolution of approx. $\sigma_E/E
= 0.20/\sqrt{\EGeV}$, considering the characteristic ratio of charged
and neutral particles in a jet. The given limit is obtained if the
\PFA\, is able to disentangle all charged from close-by neutral
showers. Since this is not possible in a realistic \PFA\, the
performance degrades due to this {\it confusion}. Any \PFA\,~relies
strongly on pattern recognition in the highly granular calorimeters
and it is not possible to distinguish between pure detector and
algorithmic effects on the reconstructed jet energy. Hence, for
reliable detector optimisation studies using a \PFA\, it is necessary
to study different \PFA\, and compare their results.

\renewcommand{\headrulewidth}{0.3pt}
\section{The Track-Based Particle Flow Algorithm}
\label{sec:TrackBasedPFA}
The Track-Based~\PFA\, is a new proposal of a \PFA\, at the
\ILC. Basis of this \PFA\, is a collection of tracks. Sequentially,
the tracks are extrapolated into the calorimeter and correlated energy
depositions are assigned to the track. Related MIP-like track segments
are identified as well. Additionally, a collection of photon
candidates can be used as an input to improve the performance of the
reconstruction. As soon as all tracks have been extrapolated their
assigned hits are removed from the collection of calorimeter
hits. Afterwards, a clustering procedure is applied on the remaining
hits to reconstruct neutral particles. A simple particle
identification~(PID) is done for charged and neutral particles. The
Track-Based~\PFA\, is implemented in~C++ within the
\MARLIN~\cite{ilcsoftportalwww,Gaede:2006pj} framework. Events for the
reconstruction are created with \MOKKA~\cite{ilcsoftportalwww}, a
\GEANT~\cite{geant4} simulation of the Large Detector Concept
(\LDC)~\cite{ldcwww}. \LCIO~\cite{Gaede:2003ip} serves as persistent
data format. The Track-Based~\PFA\, consists of six main stages:

\smallskip
\noindent{\bf{i) Photon Finding:}} Photon finding is done with the
``PhotonFinderKit'' proposed by~\cite{PhotonReconstruction}. Only ECAL
hits are taken into account. The output of this stage is a collection
of clusters labeled as photon candidates.

\smallskip
\noindent{\bf{ii) Tracking and Track-Extrapolation:}} Tracks, either
provided by Monte Carlo information or by realistic
tracking~\cite{LDCTracking}, are the basis of the Track-Based
\PFA. The tracks are sequentially extrapolated into the calorimeter
using a trajectory interface. The trajectory is given by a simple
helix model at the moment, not taking into account energy loss. If
such an extrapolation traverses one of the photon candidates it is
removed from the collection of photon candidates, since it could be
the electro-magnetic core of a hadron shower or an electron.

\smallskip
\noindent{\bf{iii) MIP-Stub Finding:}} The collection of MIP-like
energy depositions along a track extrapolation is done by a simple
geometrical procedure. A system of two cylindrical tubes are assigned
to the track extrapolation, surrounding it concentrically. Calorimeter
hits in the vicinity of the track extrapolation are sorted with
respect to their path lengths on the extrapolated trajectory. Starting
from hits with small path lengths those hits are assigned to the
MIP-stub which are located within the inner cylindrical tube. Hits
beyond the outer cylindrical tube are not taken into account. As soon
as a hit located in-between both tubes is found the procedure is
stopped. The position of the last hit collected and its direction
given by the tangent on the trajectory at this point are stored as
initial parameters for the clustering procedure performed in the next
step.

\smallskip
\noindent{\bf{iv) Clustering and Cluster-Assignment:}} The Trackwise
Clustering, proposed in~\cite{Raspereza:2006kg}, has been modified to
take the start point and direction given by the MIP-stub finding into
account. Additionally, it is adapted to produce more but smaller
clusters. The center of gravity and orientation is calculated by the
inertia tensor of each cluster. The clusters are assigned to the track
by proximity and direction criteria. The track momentum is taken into
account to prevent from assigning clusters with too much energy.

\smallskip
\noindent{\bf{v) Particle Identification and Removal of
``Charged''Calorimeter Hits:}} The~PID of charged particles is done by
a cut on the fraction of energy deposited in ECAL compared to the
HCAL. It distinguishes only between electrons and charged
pions. Additionally, muons are identified if a MIP-stub has been
assigned to the track only. Afterwards, all calorimeter hits assigned
to tracks are removed from the collection of calorimeter hits.

\smallskip
\noindent{\bf{vi) Clustering and Particle Identification on
``Neutral'' Calorimeter Hits:}} The Trackwise Clustering is applied on
the remaining calorimeter hits using the direction to the interaction
point as a start direction. The PID is done in the same way as for the
charged particles assigning a PID of photons or neutral kaons.

\smallskip
\noindent All reconstructed particles are filled into a collection
assigned to the event. The Track-Based~\PFA\, described in this note
is included in the \MARLINRECO\, package~\cite{ilcsoftportalwww}.

\section{Performance}
\begin{figure}[t!]
  \epsfxsize=8cm
  \centerline{\epsfbox{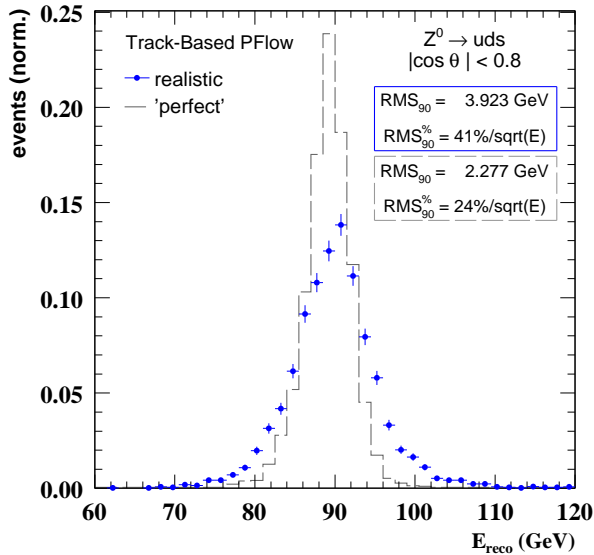}}
  \caption[Total reconstructed energy for \Zuds\, at
  $\roots=91.2$\,\GeV]{Total reconstructed energy for \Zuds\, at
$\roots=91.2$\,\GeV, realistic Track-Based~\PFA\,~(circles) and
assignment of calorimeter hits to tracks by Monte Carlo
information~(dashed lines).} \label{Fig:Reconstruction}
\end{figure}

Figure~\ref{Fig:Reconstruction} shows an example of a reconstruction
of 45~\GeV\, jets from a of \Zzero\, decaying in light-quarks~(uds) at
$\roots=91.2$\,\GeV\, using the Track-Based~\PFA~(circles). The
detector simulation has been done with
\MOKKA, using the TESLA~TDR detector
model~\cite{teslatdr,ilcsoftportalwww}. The initial direction of the
quarks is restricted to a polar acceptance of $|\cos\theta|<0.8$.
The tracks as described in stage~ii) of
Section~\ref{sec:TrackBasedPFA} are reconstructed by Monte Carlo
information. Additionally, a histogram is shown which indicates the
same reconstruction using a perfect assignment of hits to tracks by
Monte Carlo information~(dashed lines). It is getting close to the
theoretical limit of approx. $0.20 / \sqrt{\EGeV}$. The performance of
the reconstruction is measured by the root-mean-square of the smallest
range of reconstructed energies containing~90\% of the
events~($\RMS$)~\cite{Thomson:2007zz}. The Track-Based~\PFA\, reaches
a jet energy resolution of $0.41 / \sqrt{\EGeV}$ for a polar
acceptance of $|\cos\theta|<0.8$. There are two other \PFA\, available
within the \MARLIN\, framework. The first one~(\WOLF) reaches
approx. $0.52 / \sqrt{\EGeV}$~\cite{Wendt:2007iw} for the same
detector model and physics process, whereas the second one
(\PANDORAPFA) already reaches the goal of $0.30 /
\sqrt{\EGeV}$~\cite{Thomson:2007xb}.


\begin{footnotesize}

\end{footnotesize}

\end{document}